\documentstyle[11pt]{article}

\begin{document}

\begin{center}
\Large\bf
Testing  the Strong Equivalence Principle \\
with Mars Ranging Data.

\end{center}

\vskip 6pt
\begin{center}
{\sl John D. Anderson, Mark Gross,
\footnote{Department of Physics and Astronomy,
California State University,
Long Beach CA 90840.}
Eunice L. Lau,\\
Kenneth L. Nordtvedt
\footnote{Northwest Analysis, 118 Sourdough Ridge Road,
Bozeman MT 59715}
and Slava G. Turyshev
\footnote{On leave from Bogolyubov Institute for
Theoretical Microphysics, Moscow State University,

\hskip 7pt Moscow, 000958 Russia.}}
 \vskip 0.7cm
\centerline{\it Jet Propulsion Laboratory MS 301-230,}
\centerline{\it  California Institute of Technology}
\centerline{\it 4800 Oak Grove Drive - Pasadena,
CA 91109 - USA}
\end{center}
\vskip 5mm

\begin{abstract}
 The year 1996 will mark the initiation of a
number of new  space missions to the planet
Mars from which we expect to obtain a rich set of
data, including spacecraft radio tracking data.
Anticipating these events, we have analyzed the feasibility of
testing  a violation of the strong equivalence principle
({\small SEP}) with Earth-Mars ranging.
 Using analytic and numerical methods, we have demonstrated
 that  ranging data can provide a
useful estimate of the {\small SEP}  parameter $\eta$.
Two estimates of the predicted accuracy   are
quoted, one  based on conventional   covariance
 analysis, and the other based on ``modified worst case''
analysis,  which
assumes that systematic errors dominate the experiment.
If future Mars missions provide ranging measurements
with an accuracy of $\sigma$ meters, after ten years of
ranging the expected  accuracy for the  parameter
$\eta$  will be  of order $\sigma_\eta\approx (1-12)\times 10^{-4}\sigma $.
In addition,  these ranging measurements will provide a
significantly improved determination of the mass of the Jupiter system,
independent of the  test of the {\small SEP} polarization effect.

\end{abstract}
\vskip 10pt

A possible inequality of passive gravitational
and inertial masses of the same body
leads to an {\small SEP} violation, which results in observable
perturbations in the motion of celestial bodies.
Thus according to the parametrized post-Newtonian ({\small PPN}) formalism  the
ratio of
passive gravitational mass $m_g$ to inertial mass $m_i$ of a body
with rest mass $m$, may be written (Nordtvedt, 1968)
{}
$$ {m_g \over m_i} = 1  +
 \eta   {\Omega \over mc^{2}}= 1 -  \eta {G\over 2m c^{2}}\hskip 1mm
  \int \hskip -3pt \int_V   d^{3} z' d^{3}z''
{\rho (z') \rho (z'')\over |z' - z''|}, \eqno(1) $$

\noindent where {\small SEP} violation  is quantified
by the parameter $\eta$. Note that general relativity,
when analyzed in  standard {\small PPN} gauge
(Will, 1993), yields  $\eta=0$.
Whereas for  the Brans-Dicke theory,  for example,
$\eta= (2+\omega)^{-1}$, where $\omega$
is a free  dimensionless parameter of the theory.
The quantity $\Omega$ is the body's  gravitational binding energy.
The solar binding energy produces the biggest
contribution to the ratio (1) among all the celestial bodies in the solar
system.
For the standard solar model we  obtain
$\left({\Omega / mc^{2}}\right)_S \approx -3.52
\hskip1pt \cdot 10^{-6}$, which is almost four orders larger than
the Earth's  binding energy.

We maintain that   a measurement of the solar
gravitational to inertial mass ratio can be
obtained using  Mars ranging data.
In order to analyze this effect, we consider the dynamics of the
four-body Sun-Mars-Earth-Jupiter  (or {\small S-M-E-J})
system in the solar system barycentric inertial frame.
The quasi-Newtonian acceleration of   Mars  with respect
to the Sun,  $\ddot{\vec{r}}_{SM}$, is straightforwardly
calculated   to be
{}
$$ \ddot{\vec{r}}_{SM}=\ddot{\vec{r}}_M - \ddot{\vec{r}}_S =
-\mu^*_{SM} \cdot{\vec{r}_{SM} \over r_{SM}^3}
+ \mu_J \Big[ { \vec{r}_{JS} \over r_{JS}^3} - {\vec{r}_{JM} \over r_
{JM}^3}  \Big] + \eta  \Big( {\Omega \over mc^2} \Big)_{\hskip-2pt S}
\mu_J {\vec{r}_{JS} \over r_{JS}^3},  \eqno(2) $$

\noindent where  $\mu^*_{SM} \equiv \mu_S + \mu_M + \eta\Big[\mu_S
\Big( {\Omega \over mc^{2}} \Big )_{\hskip-2pt M} +
 \mu_E \Big({\Omega \over mc^2}  \Big)_{\hskip-2pt S}\Big ]$ and
 $\mu_k \equiv G m_k$.
The first and  second terms on the right side of (2) are the
classical Newtonian and  tidal acceleration terms respectively.
We denote the last term in this equation as  $\vec{a}_{\eta}$.
This is the {\small SEP} acceleration term  of order
$c^{-2}$ which may be treated as a perturbation on the
restricted three-body problem.\footnote{While it is not the only term of
that order,  the other post-Newtonian $c^{-2}$ terms
(suppressed in equation (2)) do not
affect the determination of $\eta$
until the second post-Newtonian order ($\sim c^{-4}$).}
The corresponding  {\small SEP}
 effect is evaluated as an alteration of the planetary Keplerian orbit.
 To good approximation the
{\small SEP} acceleration $\vec{a}_{\eta}$
has constant magnitude and  points in the direction from Jupiter
to the Sun, and because it  depends only on the mass
distribution in the Sun, both
Earth and  Mars experience the same perturbing acceleration.
The orbital responses of each of these planets to the   term
$\vec{a}_{\eta}$ determines the perturbation in the
Earth-Mars distance and  allows a detection of the  {\small SEP}
parameter $\eta$ by means of   ranging data.

The presence of the  acceleration term $\vec{a}_{\eta}$ in the
equations of motion  (2) results in  a
polarization of the orbits of Earth and Mars,  exemplifying the
planetary {\small SEP} effect.
By analyzing the effect of a non-zero $\eta$
on the dynamics of the Earth-Moon system moving in the
gravitational field of the Sun, Nordtvedt (1968) derived  a
polarization of the lunar orbit in the direction of the Sun
with amplitude $\delta r \sim 13 \hskip 2pt \eta $
meters (Nordtvedt effect).
The most accurate test of this effect is presently
provided by Lunar Laser Ranging ({\small LLR}),
and  in recent  results,  Dickey  {\it et al.}  (1994)
obtain
{}
$$\eta = -0.0005 \pm 0.0011 \hskip2pt.   \eqno(3)$$

\noindent
Results are also available from numerical experiments  with
combined processing of  {\small LLR}, spacecraft tracking,
planetary radar and Very Long Baseline Interferometer
({\small VLBI}) data  (Chandler {\it  et al.}, 1994).
Note that the  Sun-Mars-Earth-Jupiter system,
though governed basically by the same equations of motion as
the Sun-Earth-Moon system, is significantly different physically.
For a given value of {\small SEP} parameter $\eta$ the polarization
effects on the Earth and Mars orbits are
almost two orders of magnitude larger than on the lunar orbit.
We have examined the {\small SEP} effect on reduced
Earth-Mars ranging generated by the Deep Space Network ({\small DSN})
during  the
{\it Mariner 9} and {\it Viking} missions.
Moreover, future Mars missions,
now being planned as joint U.S.-Russian endeavours, should
yield additional ranging data.

Using both analytic and numerical methods, we have determined
the accuracy with which the   parameter $\eta$ can be measured
using Earth-Mars ranging.
The following set of the parameters were included  in a
covariance analysis;
$r_{E_0}, r_{M_0}, p_{Er_0}, p_{Mr_0}, p_{E \theta_0},
p_{M \theta_0},$   $\theta_{E_0} - \theta_{M_0}, \mu_S,
\mu_J$ and $|\vec{a}_\eta|$, where $r_{B_0}$ and $p_{Br_0}$ are the initial
barycentric  distance and corresponding momentum of planet $B$. Note  that
Jupiter's mass $\mu_J$ was taken  as  an unknown. This is because
the octopolar tide of Jupiter acting on the orbits of
Earth and Mars produces polarizations similar to
those produced by the {\small SEP} effect, but fortunately
with a different contribution  to both  orbits
and therefore   separable from the
desired {\small SEP} effect. If Jupiter's mass were uncertain by
 4 parts in $10^8$, its tidal polarization
of Mars' orbit would be uncertain by an amount equivalent
to $\eta \sim 0.001$, for example.
But Jupiter's mass is only known to about one part in a million, so one
must include  $\mu_J$ as a free parameter in analysis of the
Mars ranging experiment.
Although we performed both analytical and numerical error analyses,
we obtained the most reliable results using a
numerical integration of   (2) along with its counterpart for Earth.

While numerical integration is expected to be
more accurate than   analytic
approximations, it is possible to gain some insight into
the planetary  {\small SEP} effect by working to first order in
the eccentricity and by doing a   realistic analytical
calculation  using elliptical reference orbits for Earth and Mars.
But we found that by using  the method of
variation of parameters, we could calculate the perturbed orbits
of Earth and Mars to fourth order
in the eccentricity. This revealed that the eccentricity
correction plays a more significant role than one might
expect.  One reason for this is that the eccentricity corrections
include additional  secular terms
proportional to the time. Such  elements  dominate at large
times, and the eccentricity corrections thereby qualitatively
change the nature of the solution in the linear approximation.

In order to obtain results  for $\sigma_{\eta}$,
we assumed that $N$ daily range measurements are available from
a Mars mission, each measurement having the same
uncertainty $\sigma$ in units of meters.
The initial angles between Earth and Jupiter and Mars
and Jupiter were taken from the JPL emphemeris (DE242) at time
2441272.75, the beginning of the {\it Mariner 9} ranging measurements.
We found that the uncertainty in $\eta$ first
drops very rapidly with time and then after a few years approaches
an asymptotic behavior  $\sim N^{-1/2}$. This behavior
 gives a lower   bound on the uncertainty as
predicted  by conventional covariance analysis. For
a mission duration of   order ten years, the
uncertainty behaves as
{}
$$\sigma_{\eta} \sim 0.0039 \sigma/\sqrt{N}.   \eqno(4)$$

This result is valid for Gaussian random ranging errors
with a white spectral frequency distribution. However,
past ranging measurements using the {\it Viking Lander} have
been dominated by systematic error (Chandler {\it et al.}, 1994).
One approach to accounting for systematic error is to
multiply  the formal errors from the   covariance matrix
by $\sqrt{N}$ (Nordtvedt,  1978).  With this approach,
the expected error decreases rapidly near the beginning of the
data interval, but for large $N$ approaches an asymptotic value.
  However, we believe this is overly
conservative.  A more optimistic error estimate would include
a realistic description of the time history
of the systematic error. But a realistic systematic
error budget for ranging
data to Mars, or for   Mercury as considered by
a group at the University of Colorado (P. Bender,
private communication), is not presently available.
Yet it is unlikely
that we will be so unfortunate that the
frequency spectrum of the
signal will match the spectrum of the systematic error.
Hence we reduce the upper error bound determined by
the $\sqrt{N}$ multiplier ($\sigma_\eta = 0.0039 \sigma$)
 by a numerical factor.  The
ranging experiment proposed by the Colorado group for Mercury is
quite similar to our proposed experiment using Mars.
We therefore
follow the Colorado group and reduce the worst-case
error estimate by a factor of three and call the result the
modified worst-case analysis.  This yields an asymptotic value for
the error given by $\sigma_\eta = 0.0013 \sigma$, in our opinion
a realistic estimate of the upper error bound.

The covariance analysis gives the expected formal error in
$\mu_J$ as well. For a mission time of order ten  years
we find $\sigma_{\mu_J} \sim 5.7 \hskip 2pt  \sigma/\sqrt{N}$
in $km^3s^{-2}$, where $N$, as before, is the number of daily
ranging measurements taken during the mission.
For $\sigma = 7.9$ m, $\sigma_{\mu_J}$ falls below
the present accuracy determined from the {\it Pioneer 10} and {\it 11}
and {\it Voyager 1} and {\it2} flybys, namely  $\mu_J = 100$ $ km^3 s^{-2}$
(Campbell \& Synnott, 1985), after two years of Mars ranging.
Earth-Mars ranging can provide an impoved determination
for the mass of the Jupiter system, independent of
any determination   of the {\small SEP} effect.

For existing Mars ranging derived from the
{\it Mariner 9, Viking}, and {\it Phobos} missions, the {\sl rms} ranging
residual
referenced to the best-fit Martian orbit is 7.9 m.  We have computed
the covariance matrix with assumed daily ranging measurements for
{\it Mariner 9} (actual data interval JD 2441272.750 to JD 2441602.504)
and {\it Viking} (actual data interval JD 2442980.833 to JD
2445286.574).  Additionally,  one ranging measurement
from {\it Phobos} (actual time JD 2447605.500) was included, although it had
negligible effect on the result.  With $\sigma$ = 7.9 m,
a formal error $\sigma_\eta = 0.0012$ is obtained from the
covariance matrix. With 7.2 years of
Mars ranging, even  though not continuous, the
asymptotic limit of the modified worst-case analysis implies a
realistic error $\sigma_\eta = 0.02$, which is about 17 times the
formal error.    We conclude that the
best determination of $\eta$ is provided by the {\small LLR} data, but the
existing Mars ranging data provide an independent solar test
within  a realistic accuracy interval of
{}
$$ \sigma_{\eta} \approx 0.0012- 0.02 \hskip25pt
 ({\it Mariner \hskip 5pt 9,
\hskip 5pt Viking, \hskip 5pt  Phobos} )  \hskip10pt
 \eqno(5) $$

Future Mars Orbiter and Lander missions are expected to achieve  an {\sl rms}
systematic ranging error between 0.5 and 1.0 m.
This implies modified worst-case  realistic errors for $\eta$ and $\mu_J$ of
$\sigma_{\eta} \sim 0.0012 \sigma$ and $\sigma_{\mu_J}
\sim 1.9 \hskip 2pt  \sigma$  $km^3s^{-2}$.
Hence with mission durations of order ten years,
the interval for the uncertainties $\sigma_{\eta}$
and $\sigma_{\mu_J}$ should be
{}
$$ \sigma_{\eta} \approx \left(0.0001 - 0.0012\right)\hskip 2pt \sigma, $$
$$ \hskip 16pt \sigma_{\mu_J}\approx \left(0.09 - 1.9\right)\hskip 2pt
\sigma \hskip 10 pt {\rm km^3 s^{-2}},\eqno(6)$$

\noindent where the lower bound is based on random errors and
conventional covariance analysis, while the upper bound represents
the modified worst-case results as described in the paragraph
following (4). The expected accuracy of  future ranging
experiments should put significant constraints on
theoretical models, including a possible
inequality of the solar inertial and gravitational masses.
Although we have shown that a determination  of $\eta$ with existing {\it
Mariner 9} and {\it Viking}
ranging data is of some interest, the expected accuracy given by (5) does not
motivate us to place a high priority on a lengthy and difficult reanalysis of
those data.
Instead, we are planning on participating in future Mars missions with the goal
of generating ranging data as free of systematic error as possible,
and extending over as many years as possible.
Along the way, we most likely will obtain a test of the {\small SEP} with
existing Mars ranging. We believe William Fairbank would have encouraged us to
pursue
such an analysis, if only because we might just be surprised by the result.

We are indebted to our colleagues J. W. Armstrong
and X X Newhall  for many useful and
stimulating conversations. MG acknowledges the partial support
of an AWU-JPL sabbatical fellowship.
KLN was supported in part by the National Aeronautics and Space
Administration  through Contract NASW-4840.
SGT was supported by the  National Research Council under a
Resident Research Associateship at JPL.
JDA and ELL acknowledge that the research described in this
paper was carried out by the Jet
Propulsion Laboratory, California Institute of Technology, and
was sponsored by the Ultraviolet, Visible, and Gravitational
Astrophysics Research and Analysis Program through an agreement
with the National Aeronautics and Space Administration.

\vskip 20pt
{\Large\bf References}
\vskip 5pt
\def\ref{\vskip 5pt \par \hangindent 15pt\noindent}

\ref Campbell, J. K. \& Synnott, S. P.: 1985, {\it AJ}, {\bf90}, 364.

\ref Chandler, J. F.,  Reasenberg, R. D. \&
     Shapiro I. I.:  1994, {\it BAAS}, {\bf26},  1019.

\ref Dickey, J. O.  {\it and 11 co-authors.}:  1994,
     {\it Science}, {\bf 265},  482.

\ref Nordtvedt, K., Jr.: 1968, {\it Phys. Rev.},  {\bf169}, 1014.

\ref Nordtvedt, K., Jr.:  1978,  in: {\it A Close-Up of the Sun}.
 (JPL Pub. 78-70), 58.

\ref Will, C.M.:  1993,  {\it Theory and Experiment in Gravitational Physics.}
        (Rev. Ed.)  Cambridge University Press, Cambridge, England.

\end{document}